\documentclass[conference]{IEEEtran}

\ifCLASSINFOpdf

\else

\fi

\usepackage[cmex10]{amsmath}
\usepackage{amsfonts}
\usepackage{array,color}
\usepackage{booktabs}
\usepackage{graphicx}
\usepackage{epstopdf}
\usepackage{cite}

\begin{document}

\title{Energy Efficiency Scheme with Cellular Partition Zooming for Massive MIMO Systems}

\author{
\IEEEauthorblockN{Di Zhang\IEEEauthorrefmark{1}, Zhenyu Zhou\IEEEauthorrefmark{2}, Keping Yu\IEEEauthorrefmark{1} and Takuro Sato\IEEEauthorrefmark{1} }

\IEEEauthorblockA{\IEEEauthorrefmark{1}Globle Information and Telecommunication Studies, Waseda University, Tokyo, Japan, 169--8050\\
Email: di\_zhang@fuji.waseda.jp}
\IEEEauthorblockA{\IEEEauthorrefmark{2}State Key Laboratory of Alternate Electrical Power System with Renewable Energy Sources,\\
School of Electrical and Electronic Engineering, North China Electric Power University, Beijing, China, 102206\\
Email: zhenyu\_zhou2008@gmail.com}
}

\maketitle

\begin{abstract}
Massive multiple-input multiple-output (Massive MIMO) has been realized as a promising technology for next generation wireless mobile communications, in which Spectral efficiency (SE) and energy efficiency (EE) are two critical issues. Prior estimates have indicated that 57\% energy of the cellular system need to be supplied by the operator, especially to feed the base station (BS). While varies scheduling studies concerned on the user equipment (UE) to reduce the total energy consumption instead of BS. Fewer literatures address EE issues from a BS perspective. In this paper, an EE scheme is proposed by reducing the energy consumption of BS. The transmission model and parameters related to EE is formulated first. Afterwards, an cellular partition zooming (CPZ) scheme is proposed where the BS can zoom in to maintain the coverage area. Specifically, if no user exists in the rare area of the coverage, BS will zoom out to sleep mode to save energy. Comprehensive simulation results demonstrate that CPZ has better EE performance with negligible impact on transmission rate.
\end{abstract}

%\begin{keywords}
%Cellular networks, massive MIMO, power allocation, energy efficiency, cell coverage area.
%\end{keywords}

\IEEEpeerreviewmaketitle

\section{Introduction}
% no \IEEEPARstart
GREEN technology attacked plentiful attention worldwide because the global political agenda it is. And according to International Telecommunications Union (ITU), information communication industry (ICT) consuming 10\% energy world-wide\cite{r1}. In wireless mobile communications, it is estimated that nearly 80\% energy consumption comes from the operator while 57\% consist of cellular BS energy consumption is used to feed the BS for its coverage, QoS  of the UE transmission, etc. \cite{r2}~\cite{r3}. Hence, while talking about reducing energy consumption and enhance EE feature of cellular network, more attention should be paid to BS terminal. But unfortunately, previously, most of the studies focus on UE terminal, to reduce the energy consumption or to enhance the EE feature of the whole system by scheduling mechanisms at UE terminal or by better network deployments. Fewer attentions are paid on BS. Another, as the studies, traffic load of wireless mobile networks varies by spatial and time \cite{r4}~\cite{r5}. It is demonstrated that traffic peak epoch appears frequently in urban downtown comparing with rural village \cite{r4}. For time distribution, 10:00~12:00 together with 20:00~23:00 are traffic peak time comparing with other periods\cite{r4}~\cite{r5}. That will be realistic for energy saving while concentrating on the power allocation for various requirements and conditions.
\par
As the prior studies, EE schemes of cellular network are mainly divided into three categories, by network deployment and network architecture optimization\cite{r6}, by improving the EE feature of individual BS area\cite{r7} and by energy scheduling schemes that keep the system adaptive to variational network loads\cite{r4}~\cite{r8}.  These three categories base on different perspectives, namely network designing, BS optimization, and scheduling mechanism\cite{r1}. In\cite{r4}, a dynamic turn off scheme is proposed for energy saving. In \cite{r9}, the turn off scheme is adopted with two cellular cooperative systems. But these works only focus on traditional cellular networks conditions. In \cite{r10}, the zooming scheme is proposed towards EE of cellular networks with CoMP strategy. Where the cellular BS can zoom in to cover more area and zoom out with sleeping mode to save energy. However, the author only draws a framework without further discussion. Those days, more scholars concentrate on the BS optimization that a great deal of algorithms are proposed towards EE with BS, such as \cite{r11}~\cite{r12}. But previously seldom attention is paid to the condition in massive MIMO, which is a potential technology towards 5G\cite{r14}.  Yet in massive MIMO system, huge number of antennas are deployed. While communicating, one or more antennas that can provide better transmission quality (QoT) will be selected to serve a signal UE. Which is different from previous conditions that one antenna or 4-8 antennas working together (traditional MIMO) to serve all of the UEs. The previous zooming mechanism will not be so good while adopted in massive MIMO because the antennas with no corresponding UEs will be still working. In our previous studies\cite{r15}~\cite{r16}, some power allocation mechanisms are proposed towards massive MIMO, but antenna selection is focused in those paper. The individual BS coverage area is neglected.
\par
Based on all of those, in this paper, to further optimize the EE feature,  proposed CPZ scheme divides the coverage area of massive MIMO cellular into partition areas by radius angle and distance. That is,the coverage area is divided not only according to the distance from BS to UEs, but also the angle of coverage circumference. Then we allocate power to the fan section if there are active UEs, otherwise, turn the fan section related antennas to sleep mode. For intra-cell interference between UEs, zero forcing beamforming (ZFBF)\cite{r14} technology is adopted. Comparing with dirty paper coding (DPC)\cite{r14}, ZFBF can approach the fairly large fraction of DPC capacity. Note that in this paper, we focus on the zooming mechanism of massive MIMO cellular system to save more energy while satisfying the quality of service (QoS).  Detail discussion of antenna selection scheme\cite{r15}~\cite{r16} is left for further study. Here we assume that the selected antennas is optimal for transmission of active UEs. The zooming scheme is compared with previous scheme in\cite{r10}, results demonstrate that it possesses better EE feature with respect to the quality of service (QoS).
\par
Remaining parts of this paper are structured as follows. Section II introduces the detail system model, max transmission rate with signal to interference plus noise ratio (SINR) and radiated power, the wireless propagation model of BS and its coverage area. In Section III, the CPZ scheme is described. Section IV is the simulation results part. Detail comparison is drawn in this section. In Section V, we conclude this paper.

\section{analysis of system model}
\subsection{Analysis of Channle Model}

Massive MIMO has been approved as one feasible technology for 5G cellular network. In massive MIMO systems, a great deal of antenna dispositions was proposed, for instance, linear, rectangular, cylinder, etc.. Here we consider the cylinder disposition conditions for outdoor environment. In cylinder disposition, typically, there is one massive MIMO BS with hundreds of antennas (here the number denotes as N) and K UEs where $K<<N$ (typically, for a massive MIMO system, there are 100$\sim$200 antennas distributed by cylinder shape that serves 42 UEs.). For simplicity, we assume that the ergodic capacity can be formulated by perfect channel state information (CSI) through pilot of the transmitter. Furthermore, in this paper, only the downlink transmission is considered. The uplink transmission is left for further study.
\par
ZF-BF is used for interference eliminating where the ZF-BF matrix is $W = [w_1,w_2,...,w_k]$ . And it is proved to be asymptotically optimal for sum rate sense with respect to huge number of antennas. $W$ is utilized as the beamformer for the $N \times K$ channels with $ w_k$,  a $N\times 1$ vector. Here $h_k$ , channel from BS to UE is $1 \times N $ complex-Gaussian entries vector with zero mean  and unit variance. Then the total channel matrix, a $ K \times N $ complex matrix can be defined as $ H = [h_{1}^{T}, h_{2}^{T},..., h_{k}^{T}]$ . Then following the definition in~\cite{r12}, the ZF-BF matrix  can be expressed as:
\begin{equation}
W = H^{T}(H H^{T})^{-1}
\end{equation}
where $ (.)^H $ and $ (.)^{-1}$ stand for the conjugate and inverse transpose of a matrix.
While adoping ZFBF technology in massive MIMO, received signal of the \textit{k}th user can be expressed as:
\begin{equation}
y = h_{k} \frac{w_{k}}{\sqrt{\gamma}}x_{k} + h_{k} \sum_{i=1,i \neq k}^{K}\frac{w_{k}}{\sqrt{\gamma}}x_{k} +n_{k}
\end{equation}
where $x_{k}$ denotes the transmitted signal at the BS terminal of the \textit{k}th user, that can be expressed as $x_{k}= \sqrt{P_{k}}s_k$  with $P_k$, transmission power of UE \textit{k} of each UE. Detail relationships of total received power and BS transmission power can be formulated by wireless propagation model, which is shown in the following discussion in part B. And $s_k $ is the transmission signal at the BS terminal of \textit{k}th user that obeys independent Gaussian distribution that obeys complex Gaussian distribution with zero mean and uniform variance,  $n_k$ noise signal with zero mean and  $N_{0}$ variance, $\gamma$  the normalization factor of signal of the \textit{k}th user with expression $ \gamma = \lVert W \rVert_{F}^{2} / K $ . And $ \lVert . \rVert_{F}^{2} $ denotes for the matrix Frobenius norm.

\par
Suppose the downlink channels are constituted by components carriers (CCs) where the CCs are sufficient for the usage of kth users, and the maximum bandwidth each CC can carry is $B_{CCs}$. If $ SINR_k $  is taken as the signal to interference plus noise ratio of the \textit{k}th user. Received bit rate of \textit{k}th UE can be expressed as:
\begin{equation}
R_k = B_{CCs} \log_2 (1+SINR_k)
\end{equation}
As the assumption in this paper, $B_{CCS}$ is known with constant value, $ SINR_k $  should be focused for further study. For simplicity, we assume that the BS power is allocated equally to K UEs. That is, for \textit{k}th UE, the power is allocated by $ \beta = P_{BS}/K $ . Taking $\rho = \beta / N_0$ as the signal to noise (SNR), Thus, the SINR of UE k can be formulated as:
\begin{equation}
SINR_k = \frac{\rho|h_k w_k|^2} {\rho \sum_{j=1,j \neq k}^{j=K}|h_k w_j|^2+1}
\end{equation}
\par
According to equ.1, the beamforming vector can be deduced as:
\begin{equation}
w_k = \frac{h_{k}^{H}}{\|h_k\|}, ~~k = 1,...,K
\end{equation}
As described aforementioned, the ZFBF can eliminate the intra-interference. While subsituing equ. 5 into equ. 4, the $SINR_k$ will be:
\begin{equation}
SINR_k = \frac{\rho}{(H H^H)^{-1}/K}
\end{equation}
Then substituting equ.6 into equ.3, with arbitrary user selection mechanism and total power constraint to $P_{BS}$, the achievable ergodic rate of \textit{k}th UE can be described as:
\begin{equation}
\begin{aligned}
{R_k}' &= E\{B_{CCs} \log_2 (1+\frac{\rho K}{tr\{(H H^H)^{-1}\}})\}\\
       &\approx B_{CCs} \log_2 (1+\frac{\rho K}{E\{tr\{(H H^H)^{-1}\}\}})
\end{aligned}
\end{equation}

\par
According to the random matrix theory, the matrix $HH^H$ is a central Wishart matrix. suppose the number of antenna is larger than UE, then the matrix would be a $K \times K$ central Wishart matrix with N degrees of freedom, with covariance matrix $\textbf{I}$, a unitary matrix.  Then $tr\{(H H^H)^{-1}\}$ can be deduced as:
\begin{equation}
\begin{aligned}
tr\{(H H^H)^{-1}\} &= tr\{(H H^H)^{-1}HH^H(H H^H)^{-1}\}\\
                   &=tr\{W^HW\}\\
                   &=\|W\|_{F}^2
\end{aligned}
\end{equation}
According to random matrix theory, as K, M growing with a constant ratio $\alpha = M/K $, it  converges to a fixed deterministic value:
\begin{equation}
E\{\|W\|_{F}^2\} = \frac{K}{M-K}
\end{equation}
Then the max-sum ergodic of per cell can be formulated as:
\begin{equation}
\begin{aligned}
R_{sum} = K B_{CCs} \log_2 (1+\frac{\rho K}{E\{\|W\|_{F}^2\}})
\end{aligned}
\end{equation}

\subsection{Analysis of Propagation Model}
While considering about $SINR$, in the discussion above, like most of the previous studies, the propagation attenuation of power allocation is neglected\cite{r4}~\cite{r6}~\cite{r14} to keep things simple or focuse on other factors.  Yet attenuation do happen during  propagation in wireless environment. Here in the following step, we will take  attenuation during propagation into consideration while analysing $SINR$ to keep it closer to reality. According to wireless propagation model, after counting on the path loss, shadowing and multi path effect, received power can be formulated as:
\begin{equation}
P_k = \frac{G (\frac{r}{r_0})^{-\alpha}\varPsi P_{BS}}{K}
\end{equation}
where $G$ denotes for path gain at free space,  $r$ propagation distance,  $r_0$ omnidirectional antennas ( value of 1$\sim$10m for indoor and 10$\sim$100m for outdoor environment). $\alpha$ path loss exponent, $\varPsi$ random variable for slow fading effects, $P_{BS}$ radius power at BS terminal.
\par
While comparing the $\beta$ in section A and $P_k$ here, we can amend the $\beta$  and $\rho$ as:
\begin{equation}
\beta = \frac{G (\frac{r}{r_0})^{-\alpha}\varPsi P_{BS}}{K}
\end{equation}

\begin{equation}
\rho = \frac{G (\frac{r}{r_0})^{-\alpha}\varPsi P_{BS}}{KN_0}
\end{equation}
Take equ.13 into equ.10, the max-sum ergodic bit rate can be expressed as:
\begin{equation}
\begin{aligned}
R_{sum} &= K B_{CCs} \log_2 (1+\frac{ \rho K}{ E\{\lVert \mathrm{W} \rVert_{F}^{2}\}})\\
        &\approx K B_{CCs} \log_2 (1+\frac{ \rho K}{\frac{K}{M-K}})\\
        &=K B_{CCs} \log_2 (1+ \rho (M-K))\\
\end{aligned}
\end{equation}

\section{CPZ scheme for EE feature}
Assume that the active power of BS for UEs¡¯ cell search is puny, and further, here we only concern about the traffic request of internet service, for the voice request, it is assumed to be accomplished by other techniques, remote access unites (RAUs), CoMP, etc. In this paper, power consumption is the power for transmission together with the power attenuation during wireless propagation model, other power (e.g., circuit power, etc.) is not taken into account for intractability problem it is. Now the EE formula with bit/joule can be formulated as:
\begin{equation}
\eta = R_{sum}/P_{BS}
\end{equation}
Then the objective of EE optimization can be shown as:
\begin{equation}
 \begin{aligned}
& \max \sum_{i=0}^{K} B_{CCs} \log_2 (1+ \rho (M-K))\\
& \min \sum_{i=0}^{K} P_i\\
& s.t. ~~~\sum_{i=0}^{K}P_i \leq P_{BS},~~~P_i \in [0,P_{BS}]
 \end{aligned}
\end{equation}
where  $P_i$ is separately to each UE at the transmission terminal by BS. Now let¡¯s take a further look at equ. 16, for the first two function, they are increasing function of variable $P_i$. That is hard to fulfill them at the same time. But take the equ. 12, 13 into account, that will be possible to divide the coverage area into separate parts to decrease the power consumption while assuming the max bit rate of UEs. And as the discussion in section I, the circumference coverage of BS is not always fulfilled by UEs. If the angle of the distribution is taken into account, the power consumption can be further decreased by allocating power for transmission to the destination area but other areas.
\begin{table}[ht]
\caption{the partition zooming algorithm for EE in cellular networks} % title of Table
\small
\centering % used for centering table
\begin{tabular}{l} % centered columns (4 columns)

\hline\hline %inserts double horizontal lines
\textbf{Algorithm 1:} the CPZ algorithm \\ [0.5ex] % inserts table
%heading
\hline % inserts single horizontal line

\textbf{Step 1:} Initialization: divide the coverage area into \\
~~~~~~~~~~several pieces  by the distance and angle  \\
\textbf{Step 2:} When UE joins, execute the cell search and initialization, \\
~~~~~~~~~~report the location of UE \\
\textbf{Step 3:} Calculate the annular coverage,\\
~~~~~~~~~~\textbf{if}~$A_{new}^{UE} \subseteq \{A_{exist}^{UE}\}$, $A = A_{new}^{UE}$, \\
~~~~~~~~~~\textbf{else}~$A = A_{new}^{UE}$, update $\{A_{exist}^{UE}\}$ with new element $A_{new}^{UE}$\\
\textbf{Step 4:} Compare the distance,\\
~~~~~~~~~~\textbf{if}~$d_{new}^{UE} \leq \max {d_{exist}^{UE}}$, $d = \max {d_{exist}^{UE}}$,\\
 ~~~~~~~~~~\textbf{else}~$d = d_{new}^{UE}$\\
\textbf{Step 5:} Allocate the BS power by equ. 17 with new $A$ and $d$.\\
~~~~~~~~~~For the other areas without active UEs,\\
~~~~~~~~~~BS goes to sleep mode.\\ [1ex] % [1ex] adds vertical space

\hline %inserts single line
\end{tabular}
\label{table:nonlin} % is used to refer this table in the text
\end{table}
\par
Now we will give the EE scheme based on this conception. As shown in Fig. 1, the cylindrical square above the BS station stands for the massive MIMO antenna arrays. BS can zoom in to cover the UEs or out for energy saving. Firstly, we divided the coverage area into several parts by angle and distance that from UE to BS. While communicating, UE report its location information to BS firstly, then BS calculates the information with angle and distance, and compares it with information of exist UEs. Afterwards, allocate power for transmission. For instance, as shown in Fig. 1, the coverage area is divided into $3 \times 18$ parts by 3 annulus and 18 equal parts of circumference. For the two UEs right, the minimum annual will be the second annual area counting from the outermost, and here the $\theta$ stands for the angle of the separation. Afterwards, the BS allocates the requisite power to satisfy the QoS of transmission of the users, which is decided by the minimum power needed of the UEs in each region. It is known that the angle of a circle is $2 \pi $, then we can get the power allocation with function as:
\begin{equation}
P_i = \frac{ \rho \theta}{2 \pi}
\end{equation}

\begin{figure}
\begin{center}
\includegraphics[width=3in]{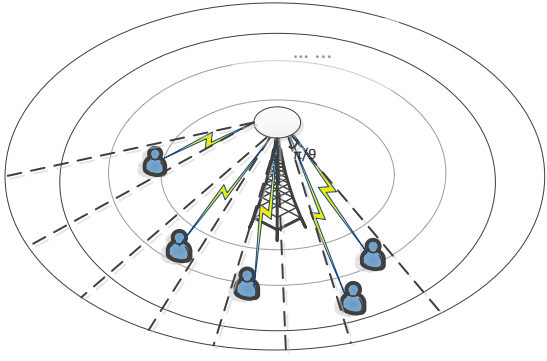}
\caption{System model with different annular region.}
\end{center}
\end{figure}

\par
This scheme can be accomplished by the LTE protocols and signaling with UEs information such as location, maximum transmission rate, etc. If we take \textit{A} stands for the angle that BS needed to cover the UE, \textit{d} distance of UEs from BS,  $d_{new}^{UE}$ distance of new UE from BS, $d_{exist}^{UE}$ distance of existed UEs from BS, $A_{new}^{UE}$ required angle of BS to cover the new UE, $A_{exist}^{UE}$ needed angle of BS to cover the existed UEs, Detail processing process can be shown as in algorithm 1.

\section{simulation results}
In this section, the proposed scheme for EE of massive MIMO system if described while comparing with traditional coverage scheme of BS and zooming scheme of BS. As for our simulation, we consider the outdoor environment, where take the value of 100 for simulation. And for the distribution of UEs in the coverage area, we assume that the max bit rates are of same value but the location for simulation. All of the values can be found in table 1.
\begin{table}[ht]
\caption{SIMULATION PARAMETERS} % title of Table
\small
\centering % used for centering table
\begin{tabular}{l l } % centered columns (4 columns)
\hline\hline %inserts double horizontal lines
Parameters & Value  \\ [0.5ex] % inserts table
%heading
\hline % inserts single horizontal line
Cell radius R & 1000m  \\ % inserting body of the table
Path loss exponent $\alpha$ & 3.7 \\
Shadow fading $\varPsi$ & 8db \\
Omnidirectional antennas r & 100m \\
Path gain G & 1 \\
Peak bit rate &20Mb/s \\
Number of antemmas M &200 \\
Bandwidth $B_{CCs}$ &5MHz \\[1ex] % [1ex] adds vertical space
\hline %inserts single line
\end{tabular}
\label{table:nonlin} % is used to refer this table in the text
\end{table}

\par
As we can see, as the bit rate has positive correlation relationship with BS power and negative correlation relationship with zoom radius, if the edge transmission is guaranteed, all of the other UEs¡¯ transmission should be guaranteed. That is because max rate transmission bit rate of UEs is set to be 20Mb/s here. And because of the distances to BS of each UE are randomly distributed, it will be meaningless for comparison with some constant number of UEs. But we can get qualitative simulation results by angle and distances of the UEs comparing with traditional BS coverage scheme.
\par
As shown in Fig.2, we compared the traditional power allocation scheme for cellular coverage with the traditional zooming scheme. In the case that without zooming, the power allocation will be always be maximum as shown by the top line in Fig. 2 for the coverage of cellular system no matter active UEs exist or not. It is clearly that the zooming scheme can reduce the power consumption as described in\cite{r10}. That is BS can zoom out to sleep mode for energy saving while no active UEs at the remote area.
\begin{figure}
\begin{center}
\includegraphics[width=3in]{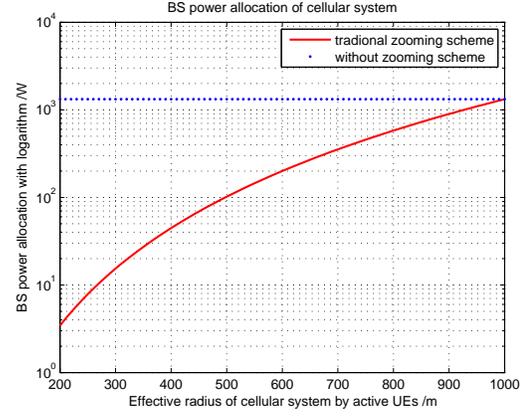}
\caption{Power allocation of cellular system with zooming scheme.}
\end{center}
\end{figure}
\par
Fig.3 compares the proposed CPZ scheme for EE of cellular system with traditional zooming scheme. With top curve denotes for the traditional zooming scheme and other remaining curves, the proposed CPZ scheme, it is clearly that proposed CPZ scheme can further reduce the energy consumption with annular partition methods. But note that the circumference divided by different factors here means that the distribution of UEs is located in the arc segment of coverage in cellular system. If other active UEs appear besides the segment coverage, the CPZ scheme should zoom in for the transmission. On condition that huge numbers of random UEs jointed randomly, the proposed CPZ scheme would approach to traditional zooming scheme.
\begin{figure}
\begin{center}
\includegraphics[width=3in]{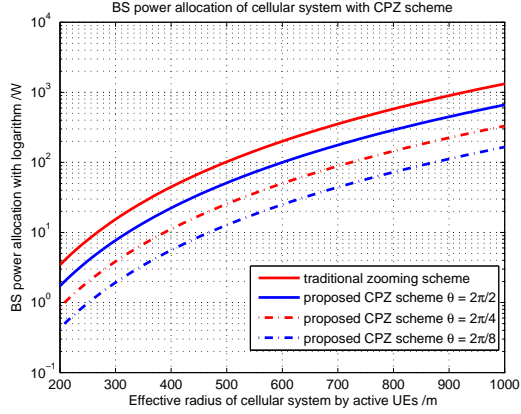}
\caption{Power allocation of cellular system with zooming scheme.}
\end{center}
\end{figure}
\par
Fig.4 manifests the energy efficiency of the proposed CPZ scheme and traditional zooming scheme. As described by the curves, the proposed CPZ scheme possess better EE character comparing with the traditional zooming scheme with further divided fan coverage area. The value of EE feature decreases with distance increasing and increases with pies that the circum is divided. It is clearly that the proposed CPZ scheme can enhance the EE feature of massive MIMO system by dividing the coverage area into separate parts with angel and distance. And as described, EE feature possesses positive correction feature with the number of segments of partition that divided.
\begin{figure}
\begin{center}
\includegraphics[width=3in]{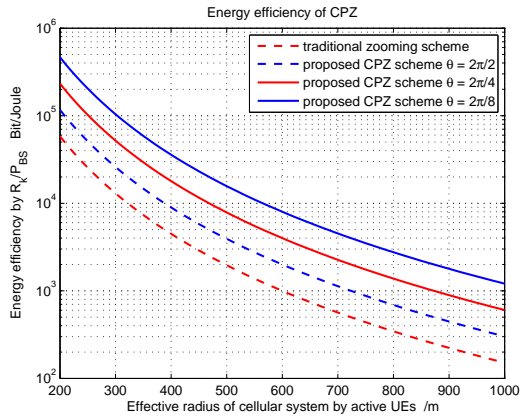}
\caption{Energy efficiency of CPZ scheme with different distances and angle.}
\end{center}
\end{figure}

\section{conclusion}
In this paper, an energy efficiency scheme for massive MIMO system has been proposed. For the coverage areas that with UEs located, BS power is allocated for transmission, in other areas, the BS goes to sleep mode for energy saving. The proposed CPZ scheme in massive MIMO cellular system can save more energy and possess better EE feature comparing with previous methods as discussed above.  And it is obviously that when the number and location distribution of UEs are given, the more segments of the annular, the more energy will be saved. The proposed CPZ scheme can be effectively applied to the cellular network for energy saving with respect to EE feature. For the huge number of UEs with random distribution, we should notice with more UEs located outside the edge area of the partition arc areas that divided, more segments should be allocated to cover the new UEs. In this case, it will approach to zooming scheme proposed previously with annular division. And if more UEs located at the remote annular coverage area, the segment division of angle would become the main factor for energy saving.  Antenna selection method will be combined next step of this study.

\section*{acknowledgement}
The author would like to thank Chinese Scholarship Council (CSC) for its financial support of this study, Yu-yang Wang, Southeast University, China, for the valuable discussion.

%\bibliographystyle{ieeetr}
%\bibliography{IEEEabrv,FirstPaper}
%\bibliography{FirstPaper}{}

\begin{bibliographystyle}{IEEEtran}
\begin{bibliography}{IEEEabrv,FirstPaper}
\end{bibliography}
\end{bibliographystyle}

% that's all folks
\end{document}